# N-ary Huffman Encoding Using High-Degree Trees - A Performance Comparison

[1]Ioannis S. Xezonakis, [2]Svoronos Leivadaros

[1]Department of Electrical and Computer Engineering, Hellenic Mediterranean University
Heraklion, 71410, Greece
xezonakis@hmu.gr

[2]Department of Electrical and Computer Engineering, Hellenic Mediterranean University
Heraklion, 71410, Greece
leivadaros@yahoo.gr

**Abstract -** In this paper we implement an *n*-ary Huffman Encoding and Decoding application using different degrees of tree structures. Our goal is to compare the performance of the algorithm in terms of compression ratio, decompression speed and weighted path length when using higher degree trees, compared to the 2-ary Huffman Code. The Huffman tree degrees that we compare are 2-ary, 3-ary, 4-ary, 5-ary, 6-ary, 7-ary, 8-ary and 16-mal. We also present the impact that branch prediction has on the performance of the *n*-ary Huffman Decoding.

**Keywords -** *n-ary Huffman Encoding, Decoding Speed, Compression Ratio, Octernary Huffman, Hexadecimal Huffman, Branch Prediction Rate*

## 1. Introduction

In 1952, David A. Huffman proposed a new lossless compression technique [1] that allowed for "the construction of minimum-redundancy codes". In this paper he described an algorithm that encoded symbols from a message to be transmitted in such a way that symbols with higher frequency of appearance were encoded by shorter length sequences than symbols with lower frequency of appearance. The proposed algorithm created uniquely decodable variable length prefix-free codes.

Huffman encoding utilizes the tree data structure to extract the code sequences for each distinct symbol that appears in the source message. In his paper, Huffman utilizes the 2-ary tree to showcase his proposed technique and offers a generalization of the process for higher degree trees ($n > 2$).

Extensive research has also been conducted on improving upon Huffman's Algorithm and implementing it in combination with other techniques to achieve higher compression ratios and encoding/decoding speeds, as well as extending its application to non-text data.

P. Suri and M. Goel [2], [3] described the implementation of the Huffman coding technique using ternary trees instead of binary, as well as a variation of it [4], the FGK Adaptive Huffman Algorithm.

R. Hashemian [5] proposes a method for clustering Huffman trees in order to achieve high efficiency in memory space as well as high-speed access to symbols.

Habib and Rahman [6] implemented the Huffman algorithm using the quaternary Huffman tree and compared the decompression rate between binary and quaternary trees; in their paper they concluded that quaternary trees offer much higher decoding speed compared to binary trees while resulting in negligibly smaller compression ratio.

In 2018, Habib et al. [7] published a paper on tribit (octernary tree) and quadbit (hexadecimal tree) Huffman encoding and compared it with previously proposed dualbit (quaternary tree) and the Zopfli algorithm [8]. They did not coincide with our experimentation results. More on this topic will be discussed in chapter 4.

In this paper we create an encoder/decoder that utilizes the *n*-ary Huffman Encoding Algorithm for $n = \{2, 3, 4, 5, 6, 7, 8, 16\}$, in order to compare the performance for the various values of *n*. We develop our application using the C programming language and focus on the encoding and decoding of text data.

A term used in the following is the branch misprediction rate. In computer architecture, a branch predictor is a digital circuit that attempts to guess whether a branch-type instruction will be taken or not. This is implemented in pipelined processor architectures in order to allow speculative execution of parts of code that normally the processor would need to wait before initiating execution. In the case that the branch predictor does not guess correctly that a branch will be taken or not taken, we call this a branch prediction miss. The ratio of branch prediction misses to the total branch predictions is called the Branch Misprediction Rate (BMR).

The effect of branch prediction on binary decision trees and specifically in Huffman trees has been noted in recent literature. However, sparse research has been conducted regarding the effects of branch prediction in decoding symbols that were encoded using the Huffman algorithm or variations of it.

In 2007, Baer presented [9] a theoretical model on the effects of static branch prediction in optimal alphabetic binary trees on microprocessor architectures such as the ARM CPU; in this paper an example of ONE-SHIFT Huffman Encoding [10] is used to showcase the proposed model.

In 2014, C. Jeong et al. [11] presented a method for improved branch prediction in Huffman decoding for streams of data via elimination of indirect branching instructions.

Another important term is Weighted Path Length (WPL). WPL is a characteristic of *n*-ary tree structures. Given an *n*-ary tree $\mathcal{T}$, with K leaf nodes and each leaf node an assigned weight value $w_i$ and depth $d_i$ the weighted path length is defined as the following

$$\text{WPL}(\mathcal{T}) = \sum_{i=1}^{K} w_i * d_i \qquad (1)$$

Where:
    $\mathcal{T}$ is the input tree
    $w_i$ is the weight of a leaf node i
    $d_i$ is the depth of a leaf node i
    K is the total number of leaf nodes

In the case of *n*-ary Huffman trees, the weighted value $w_i$ associated with each leaf node is the number of occurrences of a symbol associated with the corresponding leaf node in a text. This value basically expresses the number of transitions from one node to another (including both external and internal nodes in the tree) that will occur when the decoding process takes place.

The key values that we measure and compare between the different degrees of Huffman trees are the compression ratio, the decompression speed, the weighted path length of the tree structure and the branch misprediction rate of our application. Comparison of the encoding speed is not the main focus of this paper, however encoding speed measurements will also be presented.

## 2. The n-ary Huffman Encoding and Decoding Algorithms

In the following, we present the Encoding and Decoding Algorithms for an arbitrary *n*; the algorithms do not depend on *n*.

### 2.1. *n*-ary Huffman Encoding algorithm for message $\mathcal{M}$

- We create a list of the distinct symbols and the frequency of appearance of each symbol in the message $\mathcal{M}$. Each symbol corresponds to a leaf node of a tree $\mathcal{T}$, which grows up as described in the following.
- If the degree of the Huffman Encoder is greater than 2 (*n* > 2) we must use a number ($N_{pl}$) of placeholder 0-probability leaf nodes, in order for $\mathcal{T}$ to become a complete tree. This number is:
  $N_{pl} = (n-2) - ((S_{dst} + (n-3)) \mod (n-1))$   (2)
  Where:
      $N_{pl}$ is the number of placeholder nodes
      $S_{dst}$ is the number of distinct symbols in $\mathcal{M}$
      *n* is the degree of the Huffman tree
  The total number of leaf nodes of $\mathcal{T}$ is given by the sum $N_{pl} + S_{dst}$.
- Using the *n* nodes with the lowest frequency of appearance, we create a new parent node, whose children are the *n* nodes mentioned. This new parent node will be assigned a frequency equal to the sum of frequencies of its children and, according to it, takes its place in the list of symbols; so, newly created internal nodes are taken into consideration when searching for the lowest frequency nodes. We continue joining nodes into a single node until all leaf nodes have been linked to a parent node. The final created internal node is the root of the Huffman tree $\mathcal{T}$.
- As mentioned above, each internal node has a total number of *n* children nodes. For each internal node, we assign the value i-1 in binary form to each edge leading to the i-th child node. So, for a 5-ary tree structure, all internal nodes will have the value '000' assigned to the edge leading to their left-most child node, the value '001' to the second child node from left to right and so on. It's obvious that we use as many binary digits as required to represent the full range of values, in order to ensure that the symbols remain uniquely decodable. The number b of digits needed is given by:
  $$b = \text{ceil}(\log_2 n) \qquad (3)$$
  Where:

b is the number of bits needed
*n* is the degree of the Huffman tree utilized

For example, for $n = 3$, $b = \text{ceil}(1.584) = 2$.

Starting from the root node of T and following the path to each leaf node, we create the Huffman code for each symbol and consequently the Huffman table, containing every symbol and the corresponding code to it. Huffman codes leading to placeholder nodes are ignored. This way, the most probable symbols in M will be coded using fewer bits.

- We replace each symbol in $\mathcal{M}$ with its Huffman code sequence. We convert every 8 bits of gathered data to a byte and place the byte in a buffer. Special provision must be taken for the last bits gathered, since they will probably not complete a byte. In such a case, we add extra bits as padding, in order to complete a byte, and place the result in the buffer. Obviously, it is very important for the decoder to know how many extra bits we added as padding, in order to omit them during the decoding process of the final encoded symbol.

The contents of the buffer, which, due to the coding used, consist of a compressed form of the original data, is the main information to be sent to the decoder. However, redundant information is needed to be transmitted. The structure of the final encoded message E, which includes the "useful" data and the redundant information, mentioned above, constitute a file whose structure is presented in Fig. 1.

| 1 byte | 1 byte | 4 bytes | 4 bytes | 1 byte | Variable | Variable |
|---|---|---|---|---|---|---|
| Tree Degree | Extra Bits | Initial File Size | Compressed File Size | Number of Table Entries | Encoded Symbols | Huffman Table |

Fig. 1. The structure of the transmitted file.

The fields of this file are:
i. The degree *n* of $\mathcal{T}$.
ii. The number of extra bits we added at the end of the encoding.
iii. The number of bytes of the original file.
iv. The number of bytes of the compressed file.
v. The number of Huffman table entries.
vi. The encoded symbols themselves.
vii. The Huffman table.

## 2.2. *n*-ary Huffman Decoding algorithm for message $\mathcal{M}$

- Construct the *n*-ary Huffman tree from the Huffman table retrieved from the encoded message $\mathcal{E}$. *n* is retrieved too.

- Starting from the root node of the constructed Huffman tree, read every b bits of the encoded message and navigate to the appropriate child node denoted from the value read. The number of b bits that we need to read depends on the degree of Huffman tree we used when we encoded our message and is given from equation 3. We do this for all encoded symbols, except the last one.

- When we reach the last encoded symbol we remove any bits that we may have included at the end of the encoding process. Once again, this value is known since we included it in the message during encoding.

## 2.3. Example of 16-mal Huffman Encoding-Decoding

2-ary Huffman Encoding methodology is well established. We showcase an example of *n*-ary Huffman Encoding and Decoding using tree degree $n = 16$. Suppose we want to encode the message "Mississippi River".

### 2.3.1. 16-mal Huffman Encoding Example

- List all symbols appearing in the message together with their corresponding frequency, as in Table 1.

Table 1: Symbol distribution table.

| Symbol | Absolute Frequency |
|---|---|
| i | 5 |
| s | 4 |
| p | 2 |
| R | 1 |
| M | 1 |
| r | 1 |
| e | 1 |
| v | 1 |
| SPACE | 1 |
| Total | 17 |

- Starting tree degree is $n > 2$, therefore we use equation 2 to calculate the number $N_{pl}$ of the placeholder nodes needed.
  $N_{pl} = (16 - 2) - ((9 + (16 - 3)) \bmod (16 - 1)) = 7$
  So we will need 7 placeholder nodes.

- Take the 16 nodes with the lowest frequency of appearance and link them into 1 new node. (PH nodes are the placeholders). The new node's frequency value is the sum of the frequencies of its children nodes. In this example, we do this only once (only 16 leaf nodes).

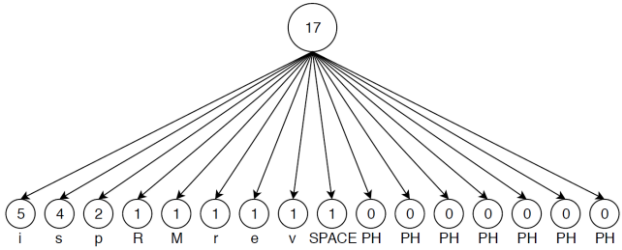

Fig. 2. Fully linked 16-mal Huffman Tree.

- Assign values to the edges of every parent node of the tree. Starting from the leftmost edge assign values 0-15 (in binary form) using for each the appropriate amount of digits. Using equation 3, this number b is:

$$b = \text{ceil}(\log_2 16) = 4$$

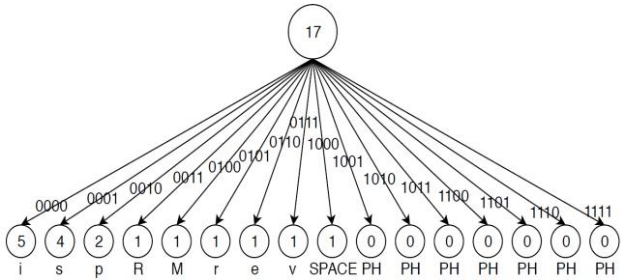

Fig. 3. 16-mal Huffman Tree with binary values assigned to edges.

- Following the path from the root node to all leaf nodes (ignoring the placeholder nodes) we get the following Huffman Table. This table is included in the final compressed and transmitted file.

Table 2: 16-mal Huffman Table for the message "Mississippi River"

| Symbol | Huffman Code |
|--------|--------------|
| i      | 0000         |
| s      | 0001         |
| p      | 0010         |
| R      | 0011         |
| M      | 0100         |
| r      | 0101         |
| e      | 0110         |
| v      | 0111         |
| SPACE  | 1000         |

- We replace each symbol with its equivalent Huffman Code, converting every 8 bits to a byte

value. Especially for the last symbol, we have 4 bits less than the necessary, in order to create the final 8-bit value, so we complete the sequence with four 0's. The number of extra added bits is included in the final transmitted file.

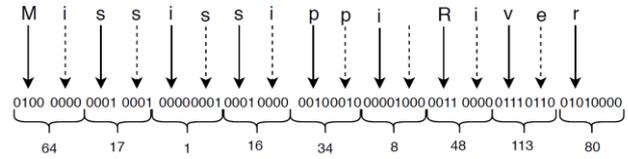

Fig. 4. Encoded message with zero-fill bits added.

Final message size is 9 bytes. Comparing it to the initial message whose size is 17 bytes, we conclude to a compression ratio of 17/9 = 1.88. Of course, since we are not using any predefined dictionaries, we must include in the final message the Huffman Table and other information about the encoding, such as the tree degree used and the extra bits added at the last encoded symbol. Practically, this means that for small text messages, Huffman Encoding without dictionaries actually yields larger files than the original.

2.3.2. 16-mal Huffman Decoding Example

- We read the Huffman table from the compressed file and construct the 16-mal tree, using it as reference. The constructed tree is not necessarily a full Huffman tree, since it won't include the placeholder nodes, but this does not influence the decompression process.
- Starting from the root node, we read the encoded message b bits at a time, navigating to the node's children from left to right. When we reach a leaf node we move back to the root node and continue reading our data. We do this for all the encoded symbols, except the last one.
- We read the number of extra bits, added during encoding of the message. In this case we added 4 bits; therefore we omit the 4 last bits of the last encoded symbol.

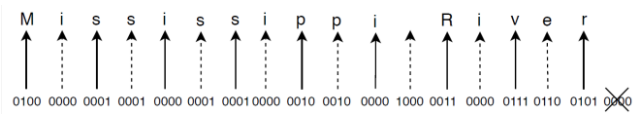

Fig. 5. Decoded Message. Last 4 bits are omitted.

## 3. System Implementation and Evaluation

3.1. Development Tools

The development platform was a Linux Ubuntu OS ver. 16.04.2 LTS computer, running on an Intel i3-3110M Quad-Core Processor clocked @ 2.4GHz. We developed the application to run in single-threaded mode. In order to measure the encoding and decoding speed we used the clock_gettime function with the MONOTONIC_CLOCK argument, which allows for measurements with nanosecond precision. Moreover, to achieve increased accuracy, we repeated the measurements 100 times when decoding, and extracted the average key values (runtime, throughput, misprediction rate).

In the following, we present measurements using our encoding/decoding application for a variety of degrees of Huffman trees, namely 2-ary, 3-ary, 4-ary, 5-ary, 6-ary, 7-ary, 8-ary and 16-mal. We present the compression speed, decompression speed, weighted path length, branch misprediction rate and compression ratio for each of the above mentioned Huffman tree degrees. Our dataset consists of a single text-only file containing randomly generated words of the English language.

During decompression, we noticed a fluctuation in performance across the different Huffman tree degrees utilized. We suspected branch misprediction to be the cause of this fluctuation, thus we used the perf Linux command utility to measure the branch misprediction rate that occurred during the execution of our decompressing application for our data set text files and for all tree degrees we implemented. We also implemented a Huffman decoder version of all tree degrees with minimal branch instructions in order to eliminate branch misprediction as a performance variable.

In all cases we used the -O3 optimization gcc argument when compiling our application. All variations were developed to run in single-threaded mode.

### 3.2. Dataset Acquisition

We used an online random word generator [12] to generate the text, consisting of 1388 Kbytes. We show in Fig. 6 the letter frequency distribution of our sample text file together with that of the letters in the English language. As expected of random word generation, the frequencies of the letters in the randomly generated text closely follow these of the English language, reflecting the real-world data distribution. The reference frequency values were obtained from [13].

In the following, various quantities are presented, according to tree degree. These quantities are compression time and compression throughput, decompression time, decompression throughput and misprediction rate, as well as compression ratio, i.e. the ratio of the original file's size to the compressed file's size.

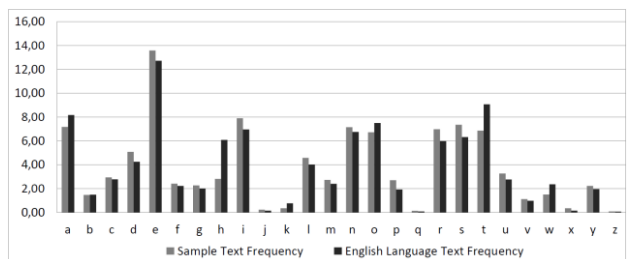

Fig. 6. Comparison of sample text data and English language characters. (Percentage frequency distribution).

### 3.3. System Benchmark Results

#### 3.3.1. Compression time and throughput

Compression time according to the tree degree utilized for the randomly generated text is presented in Fig. 7.

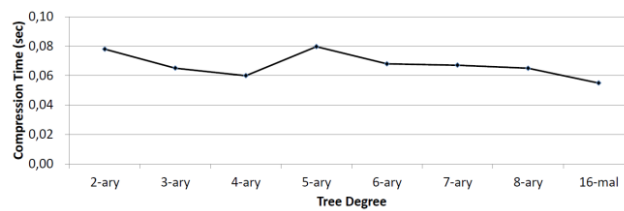

Fig. 7. Compression time of random text file (lower values are better).

As we can see, hexadecimal (16-mal) Huffman encoding offers the highest compression throughput, while quinary (5-ary) Huffman encoding offers the lowest. In Table 3, compression time appears together with compression throughput.

Table 3: Compression time and throughput of random text file.

| Tree Degree | Compression Time (sec) | Compression Throughput (Mbytes/sec) |
|---|---|---|
| 2-ary | 0.0780 | 17.37 |
| 3-ary | 0.0651 | 20.82 |
| 4-ary | 0.0600 | 22.59 |
| 5-ary | 0.0797 | 17.00 |
| 6-ary | 0.0680 | 19.93 |
| 7-ary | 0.0671 | 20.20 |
| 8-ary | 0.0650 | 20.85 |
| 16-mal | 0.0550 | 24.64 |

#### 3.3.2. Decompression time, decompression throughput and misprediction rate.

Decompression time according to the tree degree for the randomly generated text is presented in Fig. 8 while in Fig.

**9** misprediction rate according to the tree degree is presented for the same file.

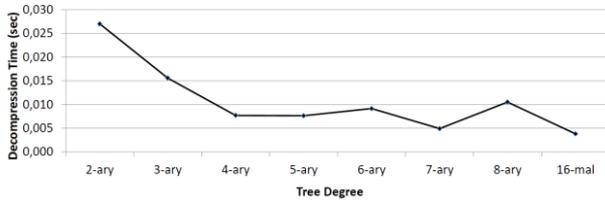

Fig. 8. Decompression time of random text file (lower values are better).

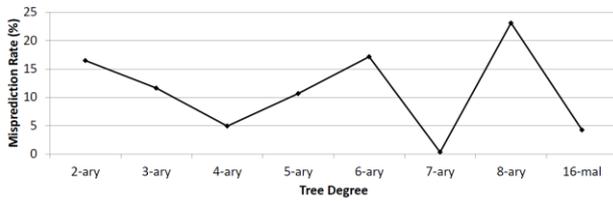

Fig. 9. Misprediction (%) rate for random text file (lower values are better).

In Table **4** and Table **5**, decompression time is presented, together with decompression throughput, misprediction rate and weighted path length. Throughput is calculated according to original file size. File I/O is not taken into account when measuring execution time

Table 4: Decompression time and decompression throughput of random text file.

| Tree Degree | Decompression Time (sec) | Decompression Throughput (Mbytes/sec) |
|---|---|---|
| 2-ary | 0.0270 | 50.22 |
| 3-ary | 0.0156 | 87.09 |
| 4-ary | 0.0077 | 176.65 |
| 5-ary | 0.0076 | 178.45 |
| 6-ary | 0.0091 | 148.90 |
| 7-ary | 0.0049 | 278.10 |
| 8-ary | 0.0105 | 129.33 |
| 16-mal | 0.0038 | 359.16 |

Table 5: Misprediction rate and weighted path length for random text file.

| Tree Degree | Misprediction Rate (%) | Weighted Path Length |
|---|---|---|
| 2-ary | 16.48 | 6271254 |
| 3-ary | 11.61 | 4037994 |
| 4-ary | 4.92 | 3201120 |
| 5-ary | 10.65 | 2783880 |
| 6-ary | 17.15 | 2533878 |
| 7-ary | 0.32 | 2369376 |
| 8-ary | 23.09 | 2251044 |
| 16-mal | 4.23 | 1668618 |

It is evident that a correlation exists between decompression times and misprediction rates for each Huffman tree structure, that is, low misprediction rate corresponds to short decompression time. Fastest decompression is achieved by 16-mal tree with 7-ary being a close second, while 2-ary offers the slowest decompression speed.

### 3.3.3. Compression Ratio

Compression ratio is the ratio of the original file's size to the compressed file's size. For example, if a file's size is 100 bytes and compressing reduces it to 40 bytes, the compression ratio is 2.50. Compression ratio is presented in Table **6** and Fig. **10** in correspondence to tree degree. Binary Huffman Encoding offers the best compression ratio while the worst one is achieved by the 5-ary Huffman Encoding.

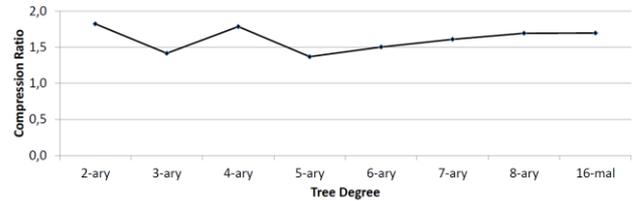

Fig. 10. Compression ratios for different Huffman tree degrees (higher values are better).

Table 6: Compression Ratio of English language text file (1388 Kbytes) for each Huffman tree degree.

| Tree Degree | Compression Ratio | Compressed Size (Kbytes) |
|---|---|---|
| 2-ary | 1.824 | 761 |
| 3-ary | 1.416 | 980 |
| 4-ary | 1.786 | 777 |
| 5-ary | 1.369 | 1014 |
| 6-ary | 1.504 | 923 |
| 7-ary | 1.610 | 862 |
| 8-ary | 1.695 | 819 |
| 16-mal | 1.697 | 818 |

### 3.3.4. Decompression measurements with minimal branch instructions

In this chapter, in Fig. 11 and Table 7, we present the performance of our modified decoding application where we removed the majority of branching instructions in order to deduce whether the branch prediction circuits of the Intel i3 processor impact the decoding performance of our application.

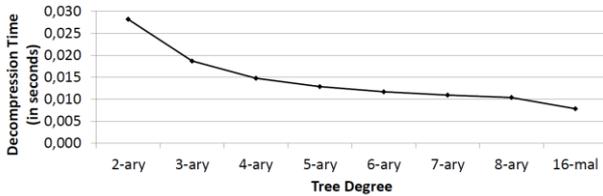

Fig. 11. Decompression time of English language text, minimal branching instructions (lower values are better).

Table 7: Decompression time, throughput and weighted path length of English language text, minimal branching instructions.

| Tree Degree | Decompression Time (sec) | Decompression Throughput (Mbytes/sec) | Weighted Path Length |
|---|---|---|---|
| 2-ary | 0.02817 | 48.12 | 6271254 |
| 3-ary | 0.01867 | 72.62 | 4037994 |
| 4-ary | 0.01474 | 91.94 | 3201120 |
| 5-ary | 0.01284 | 105.54 | 2783880 |
| 6-ary | 0.01166 | 116.24 | 2533878 |
| 7-ary | 0.01091 | 124.23 | 2369376 |
| 8-ary | 0.01037 | 130.69 | 2251044 |
| 16-mal | 0.00781 | 173.66 | 1668618 |

## 4. Results Discussion

The findings of [7] showed that utilizing 4-ary, 8-ary and 16-mal trees in Huffman Encoding yielded 60% time enhancement compared to the Zopfli Algorithm when applied to the Enwik Corpus [14] and 90% speed improvement when applied to the Canterbury Corpus [15]. This increase was shown to be more or less the same for all 3 Huffman tree degrees.

Experimental data from our research however yielded results that do not coincide with these findings. For example, we noted that as we increased the Huffman tree degree utilized, the performance of the decoder changed. In most cases, performance increased but in cases where branch misprediction rate was high it actually decreased. A more consistent, albeit slower, implementation showcases the steady increase of the decoding algorithm's performance as tree degree increases and Weighted Path Length decreases.

As seen from Fig. 10 and Table 6, the 2-ary Huffman Encoding always achieves the best compression ratio when compared to Huffman tree structures with degree greater than 2 ($n > 2$). This difference tends to be negligible when comparing Huffman Tree degrees that are powers of 2, such as 4-ary, 8-ary and 16-ary.

From Fig. 7 and Table 3 we can deduce that encoding data using 3-ary, 4-ary, 8-ary and 16-mal tree structures offer the best overall encoding throughput with mild difference in performance compared to Huffman tree degrees 2, 5, 6 and 7.

On the other hand, Table 4 shows that $n$-ary Huffman decompression performance is inconsistent. Although WPL and compression ratio dictate the total size of the encoded message and thus the time it takes to parse it when decoding it, our results come to the conclusion that WPL is not the only factor in determining decompression speed. In cases where a higher degree Huffman decompression yields much higher misprediction rate, the benefits of lower WPL will be negated by the high misprediction rate.

For example, in the case of decompressing the language text file using 8-ary Huffman Tree, we see a 115% decrease in performance compared to 7-ary when we should see a 5% increase in throughput, since weighted path length of the 8-ary Huffman tree is 5% smaller.

Estimating that branch prediction, coming from specific "if" statements in our application, may influence the performance, we created an implementation of the Huffman Decoding application that replaces "if" statements and measured the performance.

These results can be seen in Fig. 11 and Table 7. In this case decompression throughput closely follows the values of the Huffman tree's weighted path length. For example, in Table 7 the decompression runtime of the English text file using 4-ary tree is 48% shorter and its weighted path length is 49% smaller than that of the 2-ary Huffman tree. Additionally, the graph in Fig. 11 shows a more consistent increase in performance as we increase the degree of the Huffman tree used.

## 5. Conclusions

The results of this research show that the use of higher degree trees in Huffman Encoding/Decoding can be beneficial in encoding real-world text data, especially in the case of tree degrees that are powers of 2, where we achieve both high compression ratios and decompression throughputs, compared to tree degrees that are not powers of 2.

It is also shown that for specific files encoded, the Huffman tree that will be created from that file creates a decoding pattern that is favored by branch prediction circuitry in the CPU executing the decoding, thus resulting in a drastic increase in decoding performance.

Of course it comes as no surprise that branch misprediction negatively impacts performance. Rather it is noteworthy that the rate of branch misprediction shows seemingly random fluctuations across Huffman tree degrees used. We believe that future research in what leads to high misprediction rate as well as ways to consistently generate Huffman trees that yield low branch misprediction rate can lead to better future implementations of higher degree Huffman Encoding solutions.

In general however, 16-mal Huffman tree has shown to yield the best performance balance between compression throughput, decompression throughput and compression ratio, compared to other Huffman tree degrees.

**Ioannis S. Xezonakis** has been graduated in Physics (1977). He acquired MSc in Electronics (1980), MSc in Computer Automation (1989) and PhD in Informatics (1992). He has been employed in Greek Army Navy (1983-1994) and is currently a Professor of the Hellenic Mediterranean University in Software Engineering (1994 to now). His research interests cover the areas of Programming Languages, Data Structures and Algorithms Development.

**Svoronos Leivadaros** is a researcher in the fields of Computer Architecture and Intelligent Systems with experience in FPGA design and development, Data Science and Medical Information Systems. He acquired his BSc. degree from the Technological Educational Institute of Crete and his MSc. degree from the Hellenic Mediterranean University.